\documentclass[%
 aip,
 amsmath,amssymb,
 reprint,%
]{revtex4-1}

\usepackage{graphicx}
\usepackage{dcolumn}
\usepackage{bm}

\usepackage[utf8]{inputenc}
\usepackage[T1]{fontenc}
\usepackage{mathptmx}
\usepackage{newunicodechar}
\newunicodechar{⁻}{\textsuperscript{-}}
\usepackage{etoolbox}
\usepackage{ xcolor}

\usepackage{amsmath} 

\makeatletter
\def\@email#1#2{%
 \endgroup
 \patchcmd{\titleblock@produce}
  {\frontmatter@RRAPformat}
  {\frontmatter@RRAPformat{\produce@RRAP{*#1\href{mailto:#2}{#2}}}\frontmatter@RRAPformat}
  {}{}
}%
\makeatother
\begin{document}

\preprint{AIP/123-QED}

\title{\centering Hybrid Photon-Magnon Systems: Exploring the Purcell Effect}

\author{\centering\normalsize Sachin Verma}
 
\author{Abhishek Maurya}

\author{Fizaan Khan}

\author{Kuldeep Kumar Shrivastava}

\author{Rajeev Singh}

\author{Biswanath 
Bhoi}
\email{biswanath.phy@iitbhu.ac.in}

\affiliation{\centering Nano-Magnetism and Quantum Technology Lab, 
  Department of Physics, Indian Institute of Technology (Banaras Hindu University) Varanasi,
Varanasi - 221005, BHARAT (India).}

\date{\today}

\begin{abstract}
We present a novel approach to observing the Purcell effect in a photon–magnon coupled (PMC) hybrid system comprising a yttrium iron garnet (YIG) thin film and a hexagonal ring resonator (HRR) in a planar geometry. Using CST Microwave Studio, we simulated the system for varying YIG damping constants $(\alpha)$ while keeping HRR properties fixed. The results show that increasing magnon damping suppresses level-repulsion in the transmission spectra, driving the system from strong coupling into the Purcell regime via enhanced spontaneous photon emission into lossy magnons. A quantum theoretical framework accurately captures this behavior and provides estimates of the PMC strength $(g/2\pi)$, which can be tuned from 63 to 127 MHz by varying $\alpha$ from $1.4 \times 10^{-5}$ to $2.8 \times 10^{-2}$. Through simulations and parameter-space mapping, we demonstrate how damping, photon loss, and coupling strength govern the crossover between coherent coupling and the Purcell regime, while spin-density scaling confirms its cooperative nature. These findings establish a clear route for controlling photon dissipation, enabling on-chip hybrid devices and quantum technologies based on magnon–photon interactions.
\end{abstract}

\maketitle

\vspace{1cm}
\section{INTRODUCTION}
The interaction between light and matter is a fundamental aspect of quantum optics and condensed matter physics, enabling ground-breaking advancements in quantum information science and technology\cite{Wallquist2009,Kurizki2015,Harder2018,DiVincenzo2001,nielsen2010quantum,Gollwitzer2021}. One of the key phenomena governing light-matter interactions is the Purcell effect, which describes the enhancement or suppression of spontaneous emission rates of a quantum source due to its interaction with environment \cite{Stanfield2023,PhysRevA.75.053823,Krasnok2015,Zhao2023,
Gartman2023,Jordan2024,Romeira2018,Adl2020,Agarwal2024}
(as shown in Fig. 1(a)). The ability to control spontaneous emission is crucial for developing efficient single-photon sources, quantum photonic circuits, and scalable quantum information technologies \cite{Chen2024,Qian2021,Zhang2014,Zhao2023,Gollwitzer2021,Gou2024,Zhao:23}. Traditionally, the Purcell effect has been extensively studied in photon-cavity systems, where carefully engineered microcavities, photonic crystals \cite{Qian2021,David2012,Notomi2010,Akahane2003,Asano2006}, whispering gallery modes\cite{Qian2021,Kippenberg2004,Kiraz2001,Vahala2003}, and one-dimensional photonic waveguides\cite{Thyrrestrup2010} have demonstrated significant modifications of photon lifetimes and emission rates. In such systems, the Purcell factor, defined as $F \propto Q/V$ (where Q is the quality factor and V is the mode volume), dictates the enhancement of spontaneous emission, with a higher Q and a smaller V leading to stronger Purcell enhancement\cite{Qian2021}.

\begin{figure*}[htbp]
    \centering
    \includegraphics[width=\textwidth]{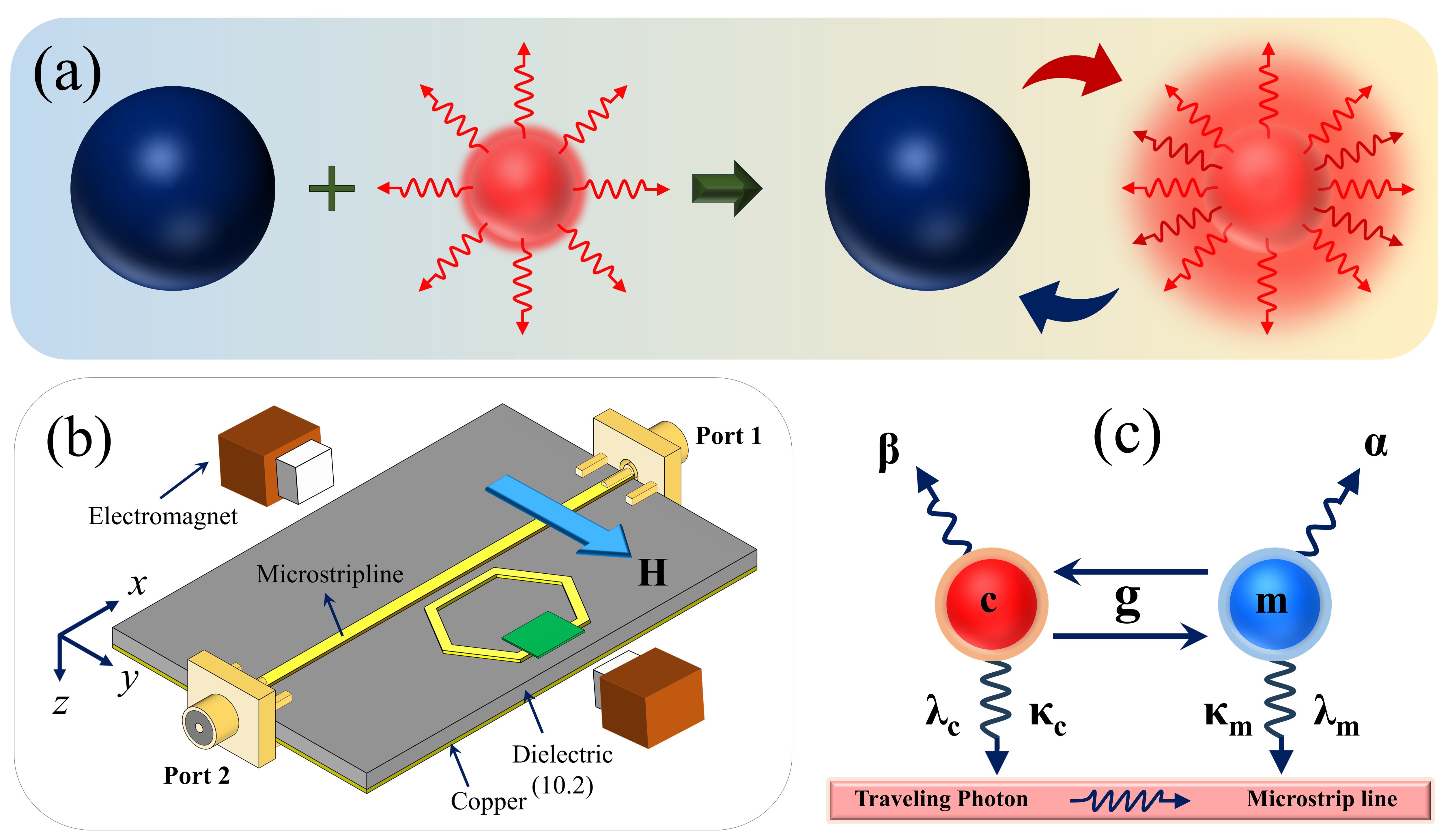}
    \caption{(a) Schematic diagram of the Purcell effect, illustrating the enhancement of the spontaneous emission rate of the quantum source mediated by the atom-field interaction. 
    (b) Simulation setup consisting of a YIG film (green) with dimensions $3 \, \text{mm} \times 3 \, \text{mm} \times 20 \, \mu\text{m}$, a microstrip line (dark yellow) with dimensions $30 \, \text{mm} \times 0.57 \, \text{mm} \times 35 \, \mu\text{m}$, a dielectric material (gray) with dimensions $30 \, \text{mm} \times 20 \, \text{mm} \times 0.64 \, \text{mm}$, and an HRR (with $a = 4 \, \text{mm}, b = 3.4 \, \text{mm}$) positioned near the microstrip line. 
    (c) Schematic of the complex photon-magnon coupled hybrid system.}
    \label{F1}
\end{figure*}

While significant progress has been made in photonic systems, the exploration of the Purcell effect in hybrid photon-magnon systems remains an emerging frontier with far-reaching implications for quantum technologies\cite{ZARERAMESHTI20221,Verma2024,Bhoi2019,Bhoi2017}. Magnons, the quanta of spin waves, offer unique advantages, including long coherence times and tunable interactions with microwave and optical photons. The coupling of magnons with photons in hybrid quantum systems\cite{Maurya2024,Kurizki2015,Wallquist2009,Clerk2020,Verma2024,Rao2020} has gained increasing attention, providing new opportunities for coherent quantum state manipulation, quantum transduction, and on-chip quantum devices\cite{Rao2019,Kiraz2001,Clerk2020,Yang2024,Androvitsaneas2016,nielsen2010quantum}. Realizing the Purcell effect in a photon-magnon coupled (PMC) system is particularly important as it enables precise control over spontaneous emission rates by leveraging both photonic and magnonic degrees of freedom. While Zhang et al.\cite{Zhang2014} provided an initial demonstration of the Purcell effect in a 3D cavity with a YIG sphere, this phenomenon remains unexplored in planar PMC hybrid systems, which are inherently compatible with chip-scale integration. Such a planar hybrid platform also offers broader tunability through extrinsic parameters (magnetic field strength and orientation) and intrinsic parameters (resonator geometry, magnon damping, and magnetic properties). Dynamically modulating magnon damping enables controlled photon dissipation, opening opportunities for tunable microwave devices, quantum memories, and scalable quantum networks.

In this work, we investigate the Purcell effect in a photon-magnon coupled planar hybrid system, focusing on the dynamic control of magnon damping to manipulate spontaneous emission rates. Unlike conventional photonic cavities, which primarily rely on optical design, our approach leverages magnetic tunability as an additional degree of freedom to enhance and control light-matter interactions. We propose a theoretical quantum model that provides deeper insights into the underlying mechanisms, validating our numerical findings and identifying optimal conditions for maximizing the Purcell effect in these systems. By tailoring magnon spin density through film thickness, we achieve direct control over the coupling strength and magnitude of Purcell enhancement. A phase diagram is further constructed to map the interplay of dissipation and coupling, revealing the regimes where the Purcell effect dominates. This combined tunability of damping and spin density creates a multidimensional control space, offering a level of control beyond conventional cavity-based approaches. The realization of the Purcell effect in PMC systems represents a significant step toward scalable and tunable quantum devices, offering a promising pathway for integrating magnonic platforms into the rapidly evolving landscape of quantum information science and technology.

\section{SIMULATION Details}
In order to explore the magnon dissipation-induced Purcell effect, we design a PMC hybrid system consisting of a hexagonal ring resonator (HRR) and a yttrium iron garnet (YIG) thin film in a planar geometry as schematically shown in Figure 1(b). The electromagnetic full-wave solver CST Microwave Studio was employed to conduct simulation and examine the dynamics of interaction between the YIG’s magnon mode and HRR’s photon mode. For microwave excitations in both the HRR and the YIG film, a microstrip feeding line was placed on the front side while the ground plane was on the back side of the sample. The dimensions of the HRR, the microstrip line, and the YIG film are indicated in the captions of Fig. 1 (b) (for further details, see Ref.\cite{Verma2024}).

For strong photon-magnon interaction, YIG film is placed directly at one edge of the HRR. We examine the transmission spectra |$S_{21}$| as a function of frequency by systematically varying the strength of the static magnetic field $H_{dc}$ applied along the y-direction. This allowed us to investigate the individual behaviors of the HRR and YIG systems, as well as their collective behavior in the hybrid configuration. Initially, a simulation was conducted for the HRR without the YIG film, revealing the pure photon mode resonance frequency $\omega_c$ peak at 5.33 GHz in the |$S_{21}$| spectrum with an associated damping constant of $\beta = \frac{\Delta\omega_{HWHM}}{\omega_c} = \frac{K_{c}}{\omega_c}$  is $4.69 \times 10^{-5}$, where linewidth $K_c/2\pi = 24.997 \, \text{MHz}$. The peak associated with the photon mode does not move with the $H_\text{dc}$ (see supplementary materials). Also, the simulations were performed for the YIG film (i.e., without the HRR), and the resonance frequency ($\omega_m$) peak position varies with the magnitude of $H_\text{dc}$ because the intrinsic precessional motion of magnetization depends on the strength of the applied field (see supplementary materials). 
\begin{figure*}[htbp]
    \centering
    \includegraphics[width=\textwidth]{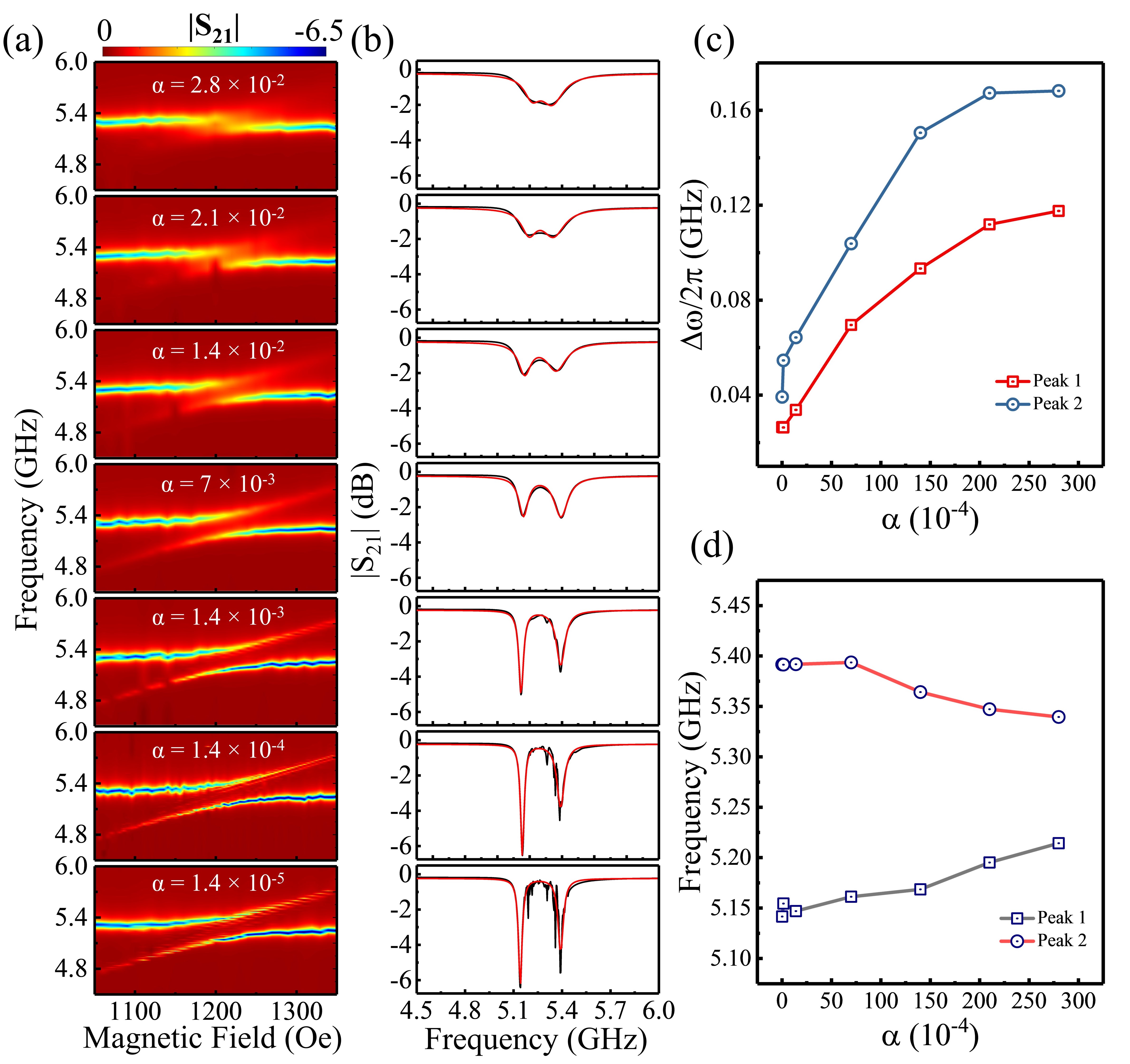}
    \caption{From the simulation results: (a) The scanned $[f, H]$ dispersion of the microwave $|S_{21}|$ spectra, showing a series of level repulsions simulated for various values of $\alpha$, and here $|S_{21}|$ is expressed in dB. (b) Transmission spectra $|S_{21}|$ as a function of microwave frequency at resonance $(\omega_c = \omega_m)$ for different values of $\alpha$, where the black curves represent simulation data and the red curves represent theoretical fitting. (c) The corresponding linewidths of the peaks. (d) The positions of the peaks.}
    \label{F1}
\end{figure*}
The frequency linewidth (HWHM) of YIG’s magnon mode was estimated as $K_m = \Delta f / 2$ and is related to the damping constant as $\alpha = K_m / \omega_m$, where $\Delta f $ is full width at half maximum, of the resonance linewidth. Subsequently, simulations including both the HRR and the YIG film were conducted in the presence of a static external magnetic field $H_\text{dc}$. Additionally, to explore the impact of magnon dissipation rates on photon-magnon interactions, simulations were also conducted by varying the damping of the YIG film from $1.4 \times 10^{-5}$ to $2.8 \times 10^{-2}$, achieved by controlling the ferromagnetic resonance linewidth ($\Delta H$).

\section{Results and Discussions}

Figure 2(a) presents the $|S_{21}|$ spectra on the $f$–$H_\text{dc}$ plane for different damping parameters of YIG, where the color variance scale represents the spectral intensity. A reduction in intensity with increasing $\alpha$ is clearly visible, while the extracted peak positions in Fig. 2(d) highlight the gradual narrowing of the split gap. Strong level-repulsion dispersion is observed for all $\alpha$ values, confirming robust coupling between the magnon mode of the YIG film and the photon mode of the HRR. However, the split gap shrinks with increasing $\alpha$, indicating a weakening of magnon–photon coupling. This reduction is accompanied by a broadening of the photon linewidth, as shown in Fig. 2(b), where the $|S_{21}|$ spectra are plotted at the resonance condition ($\omega_c = \omega_m$) for various $\alpha$. The simulation results (black curve) reveal that with higher damping, the coupled peaks shift closer and diminish in intensity. At $\alpha = 7 \times 10^{-3}$, the peaks begin to merge, and by $\alpha = 2.8 \times 10^{-2}$, they nearly collapse into a single resonance. This marks the onset of the Purcell regime, where enhanced photon dissipation dominates due to lossy magnons. In this regime, magnon damping governs the interaction dynamics, driving a crossover from strong coupling to Purcell-enhanced photon decay.

To quantify this crossover, the full-width at half-maximum (FWHM, ($\Delta\omega/2\pi$)) and central frequencies of the coupled modes are extracted from Fig. 2(b) and summarized in Figs. 2(c) and 2(d). The results reveal a parabolic trend in FWHM and a convergence of central frequencies with increasing $\alpha$, clearly demonstrating the competition between magnon damping and photon dissipation. This systematic evolution sets the stage for identifying the optimal conditions for maximizing the Purcell effect in cavity magnonics, which we discuss in the following section.

\subsection{Optimal Conditions for Purcell Effect in Cavity Magnonics}
To confirm the occurrence of the Purcell phenomenon, we analyze the coupling conditions within the coupling region. If the resonance frequencies exhibit anti-crossing and the linewidths cross at the coupling center, the coupling strength $g$ must satisfy $(K_m - K_c)/2 < g \leq K_m$\cite{Zhao:23}. On the other hand, if the dispersion shows crossing and the linewidths are anti-crossing, the condition becomes $(K_m - K_c)/2 \geq g > K_m$\cite{Zhao:23}. Since, in our case, the dispersion spectra exhibit anti-crossing (see Fig. 2a), it is essential to satisfy the above Purcell condition that $K_m$ must be greater than or equal to the coupling strength. 

From Fig. 2(b), we extracted the gap between two peak positions for all values of $\alpha$, which corresponds to the coupling strength between the HRR photon mode and the YIG magnon mode. For $\alpha = 2.1 \times 10^{-2}$ and $2.8 \times 10^{-2}$, the extracted $K_m/2\pi$ values are 111.93 MHz and 149.24 MHz, which exceed the corresponding $g/2\pi$ values of 76 MHz and 63 MHz, respectively. However, in the low damping region ($1.4 \times 10^{-5} \leq \alpha \leq 1.4 \times 10^{-2}$), $K_m/2\pi$, remains lower than the coupling strength (Table 1), and the Purcell condition is not satisfied.

Thus, the criterion for the Purcell regime is fulfilled only at higher magnon damping values, where enhanced photon dissipation dominates the hybrid dynamics. This demonstrates that the transition from strong coupling to Purcell-enhanced dissipation is not universal but depends sensitively on the interplay of magnon damping and coupling strength. To further establish this regime, it is essential to evaluate the photon dissipation rate $K_c/2\pi$, since the competition between $K_c$, $K_m$ and $g$ ultimately determines the operating regime.

\subsection{Theoretical Model and Input-Output Theory}
To quantitatively analyze variations in the photon dissipation rate ($K_c/2\pi$) and coupling strength of the PMC system, we construct a theoretical model based on the quantum harmonic oscillator using the input–output formalism\cite{Rao2020,Shrivastava2024}. A schematic representation of the quantum model, consisting of a magnon mode ($\hat{m}$) in the YIG and a microwave photon mode ($\hat{c}$) in the HRR, is illustrated in Fig. 1(c). The magnon mode is coupled to the microwave photon mode via the microstrip feeding line, which facilitates the traveling wave interaction. The total Hamiltonian for the traveling wave-mediated PMC system is  
\begin{widetext}
\begin{equation}
H = H_0 + H_p + H_\text{cmp},\tag{1a}
\end{equation}
    \begin{align} 
    H &=\hbar \tilde{\omega}_c \hat{c}^\dagger \hat{c} + \hbar \tilde{\omega}_m \hat{m}^\dagger \hat{m} + \hbar g (\hat{c}^\dagger + \hat{c})(\hat{m}^\dagger + \hat{m})  + \hbar \int \omega_k \hat{p}_k^\dagger \hat{p}_k \, \mathrm{d}k + \hbar \int \lambda_c (\hat{c} + \hat{c}^\dagger)(\hat{p}_k + \hat{p}_k^\dagger) \, \mathrm{d}k + \hbar \int \lambda_m (\hat{m} + \hat{m}^\dagger)(\hat{p}_k + \hat{p}_k^\dagger) \, \mathrm{d}k , \tag{1b}
 \end{align} 
\end{widetext}

where $H_0$ is the magnon-photon Hamiltonian, which includes the kinetic terms for both the photon and magnon modes, along with all quadratic interactions. Here, $\hat{c}^\dagger$ ($\hat{c}$) and $\hat{m}^\dagger$ ($\hat{m}$) are the respective creation (annihilation) operators of the photon and magnon modes, which have corresponding complex frequencies $\tilde{\omega}_c$ and $\tilde{\omega}_m$, respectively, and $g$ is the coupling strength between the photon and magnon resonances. The Hamiltonian $H_p$ represents the kinetic term of the traveling photon propagating through the microstrip line connected to the input and output ports, which is an integral of the wave vector over the whole real domain $(-\infty \ \text{to} \ \infty)$. Here $\hat{p}_k^\dagger$ ($\hat{p}_k$) is the boson creation (annihilation) operator of the traveling photon with $[\hat{p}_k, \hat{p}_{k'}^\dagger] = \delta(k-k')$, where $\omega_k$ is the frequency of the traveling photon with a wave vector of k. Meanwhile, the Hamiltonian $H_\text{cmp}$ describes the interaction of the traveling photon with the HRR's photon mode and the YIG's magnon mode, characterized by coupling strengths $\lambda_c$ and $\lambda_m$, respectively.

After applying the rotating wave approximation to Eq. 1 (see the Supplementary Materials), the total Hamiltonian $H$ for the coupled system becomes:
\begin{widetext}
    \begin{align} 
    H &=\hbar \tilde{\omega}_c \hat{c}^\dagger \hat{c} + \hbar \tilde{\omega}_m \hat{m}^\dagger \hat{m} + \hbar g (\hat{c} \hat{m}^\dagger + \hat{c}^\dagger \hat{m})+ \hbar \int \omega_k \hat{p}_k^\dagger \hat{p}_k \, \mathrm{d}k 
    + \hbar \int \lambda_c (\hat{c} \hat{p}_k^\dagger + \hat{c}^\dagger \hat{p}_k) \, \mathrm{d}k 
   +\hbar \int \lambda_m (\hat{m} \hat{p}_k^\dagger + \hat{m}^\dagger \hat{p}_k) \, \mathrm{d}k , \tag{2}
 \end{align} 
\end{widetext}

Here, $\tilde{\omega}_c = \omega_c - i\beta \omega_c$, and $\tilde{\omega}_m = \omega_m - i\alpha \omega_m$, where $\beta$ and $\alpha$ denote the intrinsic damping rates of the photon and magnon modes, respectively.

By solving equation (2), the equation of motion for $\hat{p}_k$ (traveling photon) can be written as
\begin{equation}
\dot{\hat{p}}_k = -\frac{i}{\hbar}\left[\hat{p}_k , H\right] = -i\omega_k \hat{p}_k - i\lambda_c\hat{c} - i\lambda_m\hat{m}\tag{3}
\end{equation}
which, after integration, gives
\begin{equation}
\hat{p}_k(t) = e^{-i\omega_k (t - t_0)} \hat{p}_k(t_0) 
- \int_{t_0}^{t} i \big[ \lambda_c  \hat{c} + \lambda_m \hat{m} \big]e^{-i\omega_k (t - t')} dt' \tag{4a}
\end{equation}
where $\hat{p}_k(t_0)$ is the initial state of operator $\hat{p}_k$ at the initial time of $t_0$, with the condition$ (t_0 < t)$. The input field operator is at the input port is defined as
\begin{equation}
\hat{p}_{in}(t) = \frac{1}{\sqrt{2\pi}}\int e^{-i\omega_k (t - t_0)} \hat{p}_k(t_0)\,dk \tag{4b}
\end{equation}

The dynamics of the photon and magnon modes can be analyzed using the master-equation formalism. Alternatively, the system may be described by deriving the corresponding quantum Langevin equations. Using the commutation relation of the creation and annihilation operators $[\hat{c}_i, \hat{c}_j^\dagger] = \delta_{ij}$, and $[\hat{c}_i, \hat{c}_j] = [\hat{c}_i^\dagger, \hat{c}_j^\dagger] = 0$, the equation of motion for the photon and magnon modes can be written as:
\begin{subequations}\label{eq:5}
\begin{align}
\dot{\hat{c}} &= -\frac{i}{\hbar}\left[\hat{c}, H\right] 
= -i\tilde{\omega}_c\, \hat{c} - ig\,\hat{m} - i\int \lambda_c \,\hat{p}_k \, dk, \label{eq:4a} \tag{5a}\\[2pt]
\dot{\hat{m}} &= -\frac{i}{\hbar}\left[\hat{m}, H\right] 
= -i\tilde{\omega}_m\, \hat{m} - ig\,\hat{c} - i\int \lambda_m \,\hat{p}_k \, dk. \label{eq:4b} \tag{5b}
\end{align}
\end{subequations}

Under the first Markov approximation, the parameters $\lambda_{c,m}$ can be regarded as constants, allowing them to be taken outside the $k$-integration. By substituting Eq. (4a) into Eq. (5), we obtained the quantum Langevin equations for the two modes, which are
\begin{subequations}\label{eq:6}
\begin{align}
\dot{\hat{c}} &= -i\tilde{\omega}_c\, \hat{c} - ig\,\hat{m} - i\sqrt{2\pi}\,\lambda_c\,\hat{p}_{in} - 2\pi{\lambda_c}^2\hat{c} - 2\pi\lambda_c\lambda_m\hat{m},  \tag{6a}\\[2pt]
\dot{\hat{m}} &= -i\tilde{\omega}_m\, \hat{m} - ig\,\hat{c} - i\sqrt{2\pi}\,\lambda_m\,\hat{p}_{in} - 2\pi{\lambda_m}^2\hat{c} - 2\pi\lambda_c\lambda_m\hat{c} \tag{6b}
\end{align}
\end{subequations}

Now, these equations can be rewritten as:
\begin{equation}
\frac{\mathrm{d}}{\mathrm{d}t} \begin{bmatrix} \hat{c} \\ \hat{m} \end{bmatrix} = -i \begin{bmatrix} \tilde{\omega}_c - i\gamma_c & g - i\sqrt{\gamma_c \gamma_m} \\ g - i\sqrt{\gamma_c \gamma_m} & \tilde{\omega}_m - i\gamma_m \end{bmatrix} \begin{bmatrix} \hat{c} \\ \hat{m} \end{bmatrix} - i \begin{bmatrix} \sqrt{\gamma_c} \\ \sqrt{\gamma_m} \end{bmatrix} \hat{p}_\text{in}(t). \tag{7a}
\end{equation}

where, $\gamma_{c,m} = 2\pi \lambda_{c,m}^2$, denotes the extrinsic damping rates of the HRR and YIG, respectively.
\begin{equation}
\frac{\mathrm{d}}{\mathrm{d}t} \begin{bmatrix} \hat{c} \\ \hat{m} \end{bmatrix} = -i H_{eff} \begin{bmatrix} \hat{c} \\ \hat{m} \end{bmatrix} - i \begin{bmatrix} \sqrt{\gamma_c} \\ \sqrt{\gamma_m} \end{bmatrix} \hat{p}_\text{in}(t). \tag{7b}
\end{equation}
The effective Hamiltonian for the coupled system becomes:
\begin{equation}
H_\text{eff} = \begin{bmatrix} \tilde{\omega}_c - i\gamma_c & g - i\sqrt{\gamma_c \gamma_m} \\ g - i\sqrt{\gamma_c \gamma_m} & \tilde{\omega}_m - i\gamma_m \end{bmatrix} =\begin{bmatrix} \tilde{\omega}_c' & g' \\ g' & \tilde{\omega}_m' \end{bmatrix}, \tag{8}
\end{equation}

where $\tilde{\omega}_c' = \tilde{\omega}_c - i\gamma_c$, $\tilde{\omega}_m' = \tilde{\omega}_m - i\gamma_m$, and $g' = g - i\sqrt{\gamma_c \gamma_m}$. The eigenvalues of the coupling matrix can be expressed as:
\begin{equation}
\tilde{\omega}_\pm = \frac{1}{2} \left[\tilde{\omega}_c' + \tilde{\omega}_m' \pm \sqrt{\left(\tilde{\omega}_c' - \tilde{\omega}_m'\right)^2 + 4g'^2}\right] \tag{9a}
\end{equation}

We have calculated the complex eigenvalues of two coupled modes denoted as $\tilde{\omega}_\pm = \omega_\pm - i\Gamma_\pm$. Here the $\omega_\pm$ represents the real part of the eigenvalue, corresponding to the higher and lower energy modes in the dispersion curve, while $\Gamma_\pm$ is the imaginary part, which describes the linewidth evolution of the coupled modes. In the expression $g' = g - i\sqrt{\gamma_c \gamma_m}$, when the real part dominates,$\omega_+$ and $\omega_-$ both will repel each other, which is referred to as level-repulsion. Conversely, when the imaginary part dominates, the two modes will cross each other, a phenomenon known as level-attraction.

The frequency gap, $\Delta$ between the two modes at the anti-crossing center $(\omega_c = \omega_m)$ is given by $\Delta = (\tilde{\omega}_+ - \tilde{\omega}_-)/2\pi$. 
Using Equation (9a), $\Delta$ can be expressed as: 
\begin{equation}
\Delta = \frac{1}{2\pi} \left[ \sqrt{4\left(g'\right)^2-\left[\omega_c\left(\beta-\alpha\right)+ \left(\gamma_c-\gamma_m\right)\right]^2 }\right] \tag{9b}
\end{equation}
To calculate the transmission coefficient ($|S_{21}|$) of the PMC system, the steady-state equations in the frequency domain are obtained by applying a Fourier transform to Eq. (7), which gives:

\begin{align}
    &i(\omega-\tilde{\omega_c})\hat{c}(\omega)-i\sqrt{\gamma_c}\,\hat{p}_{in}(\omega)-\gamma_c\,\hat{c}(\omega)-\sqrt{\gamma_c\gamma_m}\,\hat{m}(\omega) \nonumber \\
    &-ig\hat{m}(\omega)=0 \tag{10a}
    \\
    &i(\omega-\tilde{\omega_m})\hat{m}(\omega)-i\sqrt{\gamma_m}\,\hat{p}_{in}(\omega)-\gamma_m\,\hat{m}(\omega)-\sqrt{\gamma_c\gamma_m}\,\hat{c}(\omega) \nonumber \\
    &-ig\hat{c}(\omega)=0 \tag{10b}
\end{align}

From Eq. (4a), for ${t_1}>t$, the traveling photon $\hat{p}_k$ after its interaction with the two modes is expressed as:
\begin{equation}
\hat{p}_k(t) = e^{-i\omega_k (t - t_1)} \hat{p}_k(t_1) 
- \int_{t}^{t_1} i \big[ \lambda_c  \hat{c} + \lambda_m \hat{m} \big]e^{-i\omega_k (t - t')} dt' \tag{11a}
\end{equation}

The field operator corresponding to the output port is defined as:
\begin{equation}
\hat{p}_{out}(t) = \frac{1}{\sqrt{2\pi}}\int e^{-i\omega_k (t - t_1)} \hat{p}_k(t_1)\,dk \tag{11b}
\end{equation}
By substituting Eq. (11) into Eq. (5), the time-retarded Heisenberg–Langevin equations for the coupled system can be obtained in both the time and frequency domains.
\begin{equation}
\frac{\mathrm{d}}{\mathrm{d}t} \begin{bmatrix} \hat{c} \\ \hat{m} \end{bmatrix} = -i \begin{bmatrix} \tilde{\omega}_c + i\gamma_c & g + i\sqrt{\gamma_c \gamma_m} \\ g + i\sqrt{\gamma_c \gamma_m} & \tilde{\omega}_m + i\gamma_m \end{bmatrix} \begin{bmatrix} \hat{c} \\ \hat{m} \end{bmatrix} - i \begin{bmatrix} \sqrt{\gamma_c} \\ \sqrt{\gamma_m} \end{bmatrix} \hat{p}_\text{out}(t). \tag{12}
\end{equation}

Applying a Fourier transform to Eq. (12) yields:
\begin{align}
    &i(\omega-\tilde{\omega_c})\hat{c}(\omega)-i\sqrt{\gamma_c}\,\hat{p}_{out}(\omega)+\gamma_c\,\hat{c}(\omega)+\sqrt{\gamma_c\gamma_m}\,\hat{m}(\omega) \nonumber \\
    &-ig\hat{m}(\omega)=0 \tag{13a}
    \\
    &i(\omega-\tilde{\omega_m})\hat{m}(\omega)-i\sqrt{\gamma_m}\,\hat{p}_{out}(\omega)+\gamma_m\,\hat{m}(\omega)+\sqrt{\gamma_c\gamma_m}\,\hat{c}(\omega)
    \nonumber \\
    &-ig\hat{c}(\omega)=0 \tag{13b}
\end{align}

By solving Eqs. (10) and (13), the relation between the input and output fields of this coupled system can be obtained as:
\begin{align}
    \hat{p}_{out} = \hat{p}_{in} - 2i\sqrt{\gamma_c}\,\hat{c}(\omega) - 2i\sqrt{\gamma_m}\,\hat{m}(\omega)\tag{14}
\end{align}

Solving Eqs. 10(a) and (b) provides the field operators expressed in terms of the input field as:
\begin{widetext}
\begin{align}
    \hat{c}(\omega) = \frac{\hat{p}_{in}\left[g\sqrt{\gamma_m}+i\alpha\sqrt{k_c}\,\omega_m+\sqrt{k_c}\,\omega-\sqrt{k_c}\,\omega_m \right]}{\left[(ig + \sqrt{\gamma_c \gamma_m})^2 + (i\beta \omega_c + i\gamma_c + \omega - \omega_c)(i\alpha \omega_m + i\gamma_m + \omega - \omega_m)\right]} \tag{15a}
\end{align}
\begin{align}
    \hat{m}(\omega) = \frac{\hat{p}_{in}\left[g\sqrt{\gamma_c}+i\beta\sqrt{k_m}\,\omega_c+\sqrt{k_m}\,\omega-\sqrt{k_m}\,\omega_c \right]}{\left[(ig + \sqrt{\gamma_c \gamma_m})^2 + (i\beta \omega_c + i\gamma_c + \omega - \omega_c)(i\alpha \omega_m + i\gamma_m + \omega - \omega_m)\right]} \tag{15b}
\end{align}

Therefore, the transmission coefficient of the coupled system can be expressed as:
\begin{equation}
|S_{21}| = \frac{\hat{p_{out}}}{\hat{p}_{in}}-1 \tag{16a}
\end{equation}
\begin{equation}
\|S_{21}| = \frac{2\alpha \gamma_c \omega_m + 2\beta \gamma_m \omega_c - 4i g \sqrt{\gamma_c \gamma_m} - 2i\gamma_c (\omega - \omega_m) - 2i\gamma_m (\omega - \omega_c)}{\left[(ig + \sqrt{\gamma_c \gamma_m})^2 + (i\beta \omega_c + i\gamma_c + \omega - \omega_c)(i\alpha \omega_m + i\gamma_m + \omega - \omega_m)\right]} \tag{16b}
\end{equation}
\end{widetext}

\begin{figure}
    \centering
    \includegraphics[width=\linewidth]{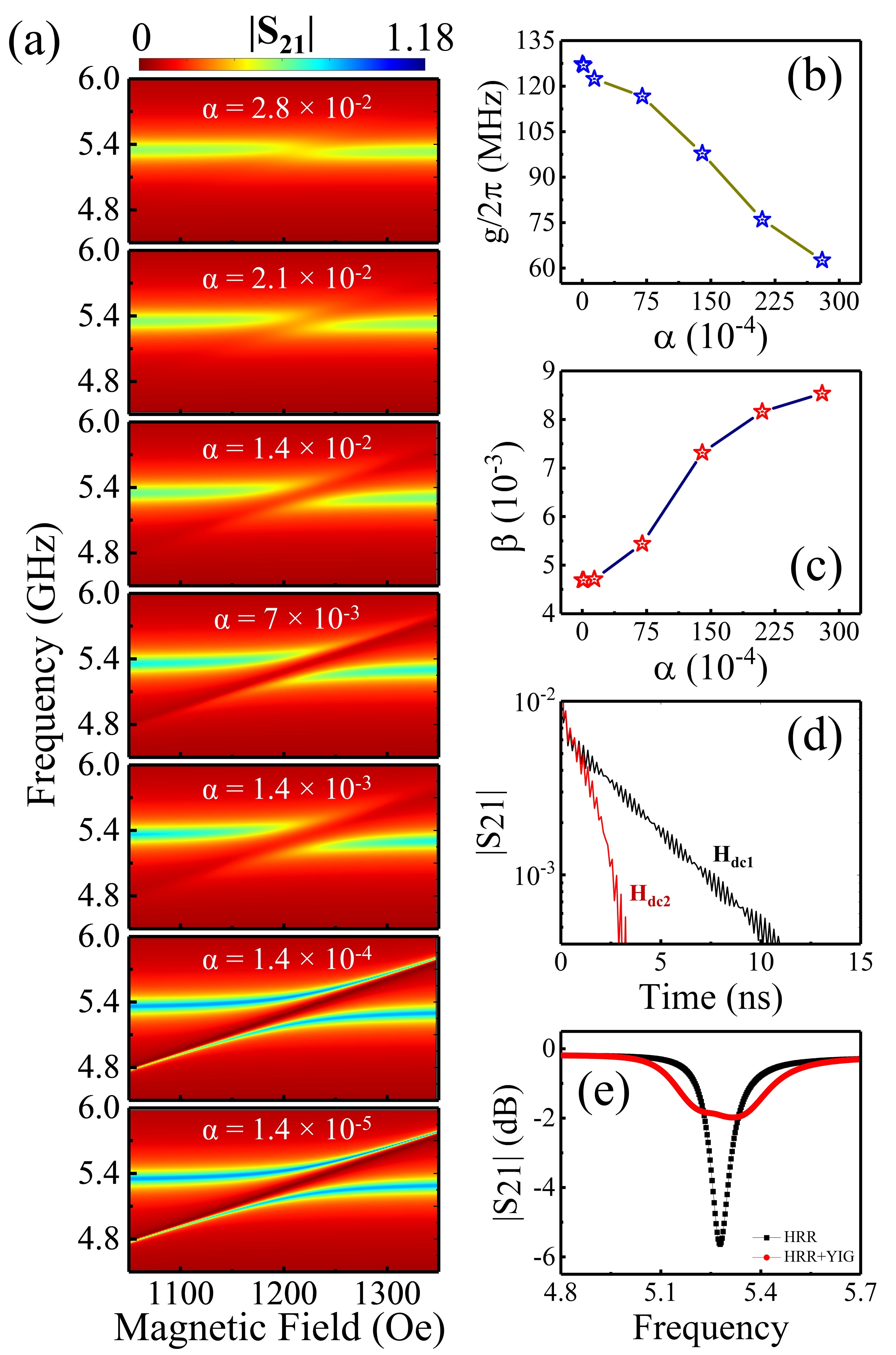}
    \caption{(a) $|S_{21}|$ transmission coefficient of photon-magnon coupled modes on $f-H$ plane, calculated using a quantum model for different values of magnon damping. The variation of (b) coupling strength $(g/2\pi)$, and (c) the damping constant of the photon mode $(\beta)$, both as functions of magnon damping $(\alpha)$. (d) The $|S_{21}|$ transmission Spectra as a function of time, are calculated using the FFT of simulation data; and (e) The $|S_{21}|$ spectra as a function of frequency, for two different bias magnetic fields ($H_{dc1} = 900 Oe$, and $H_{dc2} = 1210 Oe$) applied along the Y-direction, shown only for the highest magnon damping $(\alpha = 2.8 \times 10^{-2})$. }
    \label{F2}
\end{figure}

\begin{table*}[ht]
    \centering
    \begin{tabular}{|c|c|c|c|c|c|c|}
        \hline
        \textbf{Sr. No.} & \textbf{$\alpha$ (10$^{-4}$)} & \textbf{$K_m/2\pi$ (MHz)} & \textbf{$K_c/2\pi$ (MHz)} & \textbf{$g/2\pi$ (MHz)} & \textbf{$\beta$ (10$^{-3}$)} & \textbf{$(K_m - K_c)/2 < g \leq K_m$} \\ \hline
        1 & 0.14 & 0.07462 & 24.99 & 127.3  & 4.688  & No \\ \hline
        2 & 1.4  & 0.7462  & 24.997 & 126.9  & 4.7    & No \\ \hline
        3 & 14   & 7.462   & 25.15  & 122.38 & 4.718  & No \\ \hline
        4 & 70   & 37.31   & 29     & 116.61 & 5.44   & No \\ \hline
        5 & 140  & 74.62   & 39     & 97.82  & 7.317  & No \\ \hline
        6 & 210  & 111.93  & 43.5   & 76.03  & 8.161  & Yes \\ \hline
        7 & 280  & 149.24  & 45.5   & 62.6   & 8.536  & Yes \\ \hline
    \end{tabular}
    \caption{Comparison of the Purcell Condition for different values of magnon damping $(\alpha)$, based on the frequency linewidth (HWHM) of magnon mode $(K_m)$, frequency linewidth of photon mode $(K_c)$, coupling strength $(g)$, and damping constant of photon mode $(\beta)$.}
    \label{tab:data_table}
\end{table*}

\subsection{Numerical verification of Purcell Condition}
To further validate the analytical framework, we examine the numerical signatures of the Purcell effect by evaluating the transmission spectra using Eq. (16). Figure 3(a) shows the calculated $|S_{21}|$ transmission amplitude coefficient as a function of frequency for different magnon damping values which closely follows the simulated results in Fig. 2(a), demonstrating a good agreement between the theoretical model and the simulations. From the calculated photon mode linewidth $K_c/2\pi$, coupling strength $g/2\pi$, and magnon linewidth $K_m/2\pi$, summarized in Table 1, we find that the Purcell condition, $((K_m - K_c)/2 < g \leq K_m)$, is satisfied only for $\alpha \geq 2.1 \times 10^{-2}$. The corresponding values of $K_c/2\pi$ are 39 MHz and 45.5 MHz, while $g/2\pi$ are 76.03 MHz and 62.6 MHz, respectively. These values confirm that the Purcell effect emerges at higher magnon damping, thereby validating both the accuracy and predictive power of our theoretical model.

To further examine the influence of magnon damping on coupling dynamics, we calculated the coupling strength using Eq. (9), with the results shown in Fig. 3(b). The analysis shows that the coupling strength decreases gradually for low magnon damping, reaching 116.61 MHz, but exhibits a pronounced reduction at higher damping, dropping to 62.6 MHz for $\alpha = 2.8 \times 10^{-2}$. The photon damping rate, calculated using Eq.~(9) (see Table~1) and shown in Fig.~3(c), increases significantly with higher magnon damping and exhibits enhanced spontaneous emission only for $\alpha \geq 2.1 \times 10^{-2}$.

\begin{figure}
    \centering
    \includegraphics[width=\linewidth]{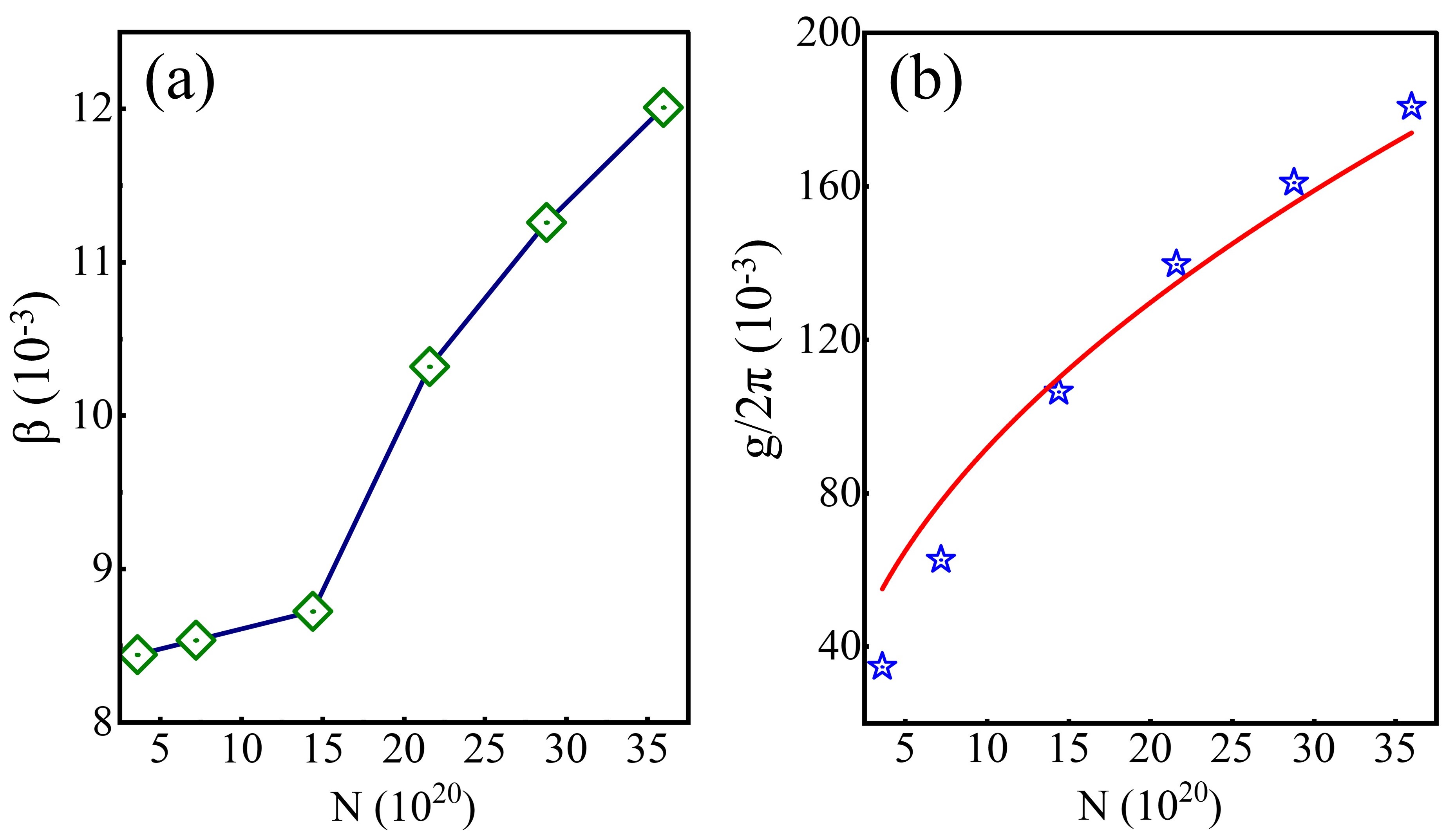}
    \caption{(a) Extracted values of $\beta$ as a function of spin density, showing systematic enhancement of magnon damping with increasing spin reservoir, consistent with the Purcell effect. (b) Calculated coupling strength $g$ versus total spin number $N$, exhibiting the expected $g = g_0 \sqrt{N}$ scaling characteristic of collective coherent interactions.}
    \label{F2}
\end{figure}

Furthermore, for the highest magnon damping ($\alpha = 2.8 \times 10^{-2}$), we calculated the FFT of the transmission spectra for two bias magnetic fields, $H_{\text{dc}_1} = 900$~Oe and $H_{\text{dc}_2} = 1210$~Oe, and plotted $|S_{21}|$ vs. time in Fig.~3(d). The figure demonstrates the enhanced photon decay rate at the coupling center (i.e., $H_{\text{dc}_2}$) compared to an uncoupled state (i.e., $H_{\text{dc}_1}$). For a clearer comparison, the transmission spectra for HRR alone and HRR+YIG are shown in Fig.~3(e). This figure highlights the increase in HRR linewidth from 24.99 MHz to 45.5 MHz due to the Purcell effect. These findings provide a comprehensive understanding of the system's transition into the Purcell regime.

\subsection{Control of Purcell Effect}
To further explore the controllability of the Purcell effect in planar PMC systems, we examined its dependence on spin density by tuning the spin volume through controlled variation of the YIG film thickness. The corresponding color maps of the transmission spectra for different spin volumes are presented in Supplementary Fig. S3(a), with theoretical simulations shown in Fig. S3(b). Using the quantum model and Eq. (9b), the values of $\beta$ were extracted as a function of spin density [Fig. 4(a)], which clearly demonstrate that the damping of the HRR increases systematically with larger spin reservoirs. This behavior signifies that the Purcell effect manifested as enhanced photon dissipation becomes more dominant when the collective spin ensemble is larger, thereby accelerating photon leakage into the magnonic channel. In parallel, the coupling strength obtained from simulations, shown in Fig. 4(b), follows the expected square-root law $g = g_0\sqrt{N}$, with $N$ being the total number of spins. This $g = g_0\sqrt{N}$ scaling is a hallmark of collective coherent interactions, confirming that the enhancement of the Purcell effect originates from cooperative spin ensemble dynamics rather than single-spin processes. These findings highlight that both magnon damping and spin density serve as tunable parameters for dynamically controlling the Purcell effect, thus offering a multidimensional design space for optimizing photon–magnon hybrid platforms in quantum technologies.

While the above analysis clarified the influence of magnon damping and spin density, a more comprehensive framework is needed to delineate the parameter regimes governing the transition from coherent coupling to the Purcell regime. To this end, we mapped the interdependence of $\alpha, \beta$, and $g$ in a unified parameter space. Since the coupling under investigation is coherent, we focus only on the real part of $\Delta$ to capture the effective mode splitting. Figure 5, obtained from Eq. (9b) with $\omega_c=5.33$ GHz, presents the phase diagram where the color gradient of $Re(\Delta)$ visualizes the effective coupling strength: red–yellow regions correspond to strong coupling, whereas blue regions indicate weak interaction. This representation clearly illustrates how increasing magnon damping systematically drives the system from strong coherent coupling toward the Purcell regime. As expected, coupling strength decreases monotonically with higher $\alpha$, with deep-blue regions denoting the weakest coupling at large $\alpha$ and red regions highlighting the strongest interaction at low $\alpha$. The black star markers, representing $\beta$ values calculated for different $\alpha$, align within the red–orange volume, confirming their placement in the cavity-enhanced decay regime. Although these points do not fully satisfy the Purcell condition, their systematic shift toward higher $\beta$ with increasing $\alpha$ outlines a clear transition pathway into the Purcell-dominated regime. Furthermore, the shaded region in Fig. 5 marks the parameter space where the Purcell condition is met, showing that the two highest damping cases lie squarely within this region, while the shaded region in Fig. S4 of the Supplementary provides a magnified view of the same parameter space where the Purcell condition is satisfied. Together, these results provide a comprehensive visualization of how $\alpha, \beta$, and $g$ dictate the system’s operational regime, establishing a clear framework for engineering transitions between coherent coupling and Purcell-enhanced dissipation in planar PMC systems.

Our results demonstrate that achieving the Purcell effect in a photon-magnon coupled hybrid system requires the use of lossy magnetic materials. The damping of YIG can be tuned over a wide range $10^{-1}$ to $10^{-5}$ through various techniques, including growth optimization, rare-earth metal doping, spin pumping, nonlocal spin currents, and substrate engineering. Growth optimization, achieved by modifying deposition and annealing conditions or using different substrates, enables intrinsic control over damping\cite{bhoi2018stress}. Rare-earth doping introduces impurity scattering, influencing relaxation dynamics\cite{gurjar2021control}, while spin pumping provides an external approach where a spin-polarized current, injected via a nanocontact, dynamically modulates the damping\cite{sun2013damping,heinrich2011spin,navabi2019control}. Additionally, nonlocal spin currents and substrate orientation adjustments fine-tune Gilbert damping, as demonstrated in prior studies\cite{gurjar2021control,haidar2016controlling}.

\begin{figure}
    \centering
    \includegraphics[width=\linewidth]{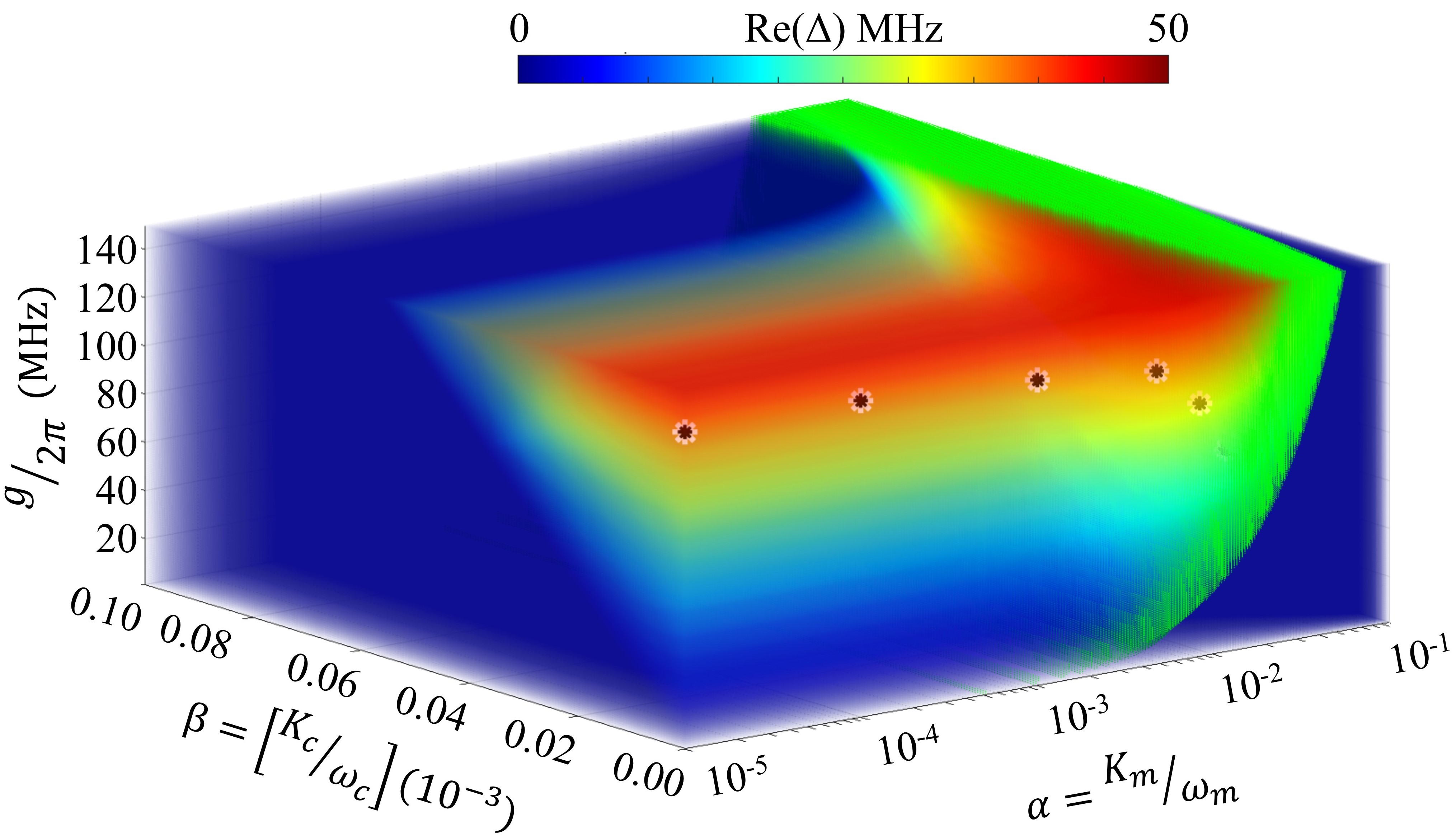}
    \caption{Analytically derived three-dimensional phase diagram of the real part of $\Delta$ depicting various forms of anti-crossing dispersion within the $\left( \alpha-\beta-g\right)$ parameter space. The color gradient represents the variation of the coupling strength $\Delta$, while the green-shaded region highlights the Purcell regime. Theoretical $\beta$ values are marked with solid black star marker with white outline.}
    \label{F2}
\end{figure} 

Temperature is another critical factor influencing ferromagnetic resonance linewidth and magnon lifetime. While our analysis has been based on room-temperature parameters most relevant for practical on-chip applications magnon damping exhibits strong temperature dependence. For instance, Jermain et al. reported a substantial increase in linewidth in thin-film YIG at low temperatures due to impurity relaxation\cite{jermain2017increased}, whereas Kosen et al. demonstrated exceptionally low linewidths (1.4 MHz) in high-quality YIG at millikelvin temperatures, dominated by coupling to two-level fluctuators \cite{kosen2019microwave}. These studies underscore that although room-temperature operation ensures technological relevance, cryogenic operation can further enhance magnon coherence and open additional opportunities for quantum devices.

A comparison between our obtained coupling strengths and previously reported values for millimetre-scale YIG magnets is summarized as follows. Harder et. al. observed a coupling strength of 31.5 MHz for a YIG sphere in a 3D cavity\cite{harder2016study}, while Hyde et al. reported 30.3 MHz for a similar YIG sphere/3D cavity system\cite{hyde2016indirect}. Kaur et al. demonstrated a larger value of 65 MHz for a YIG sphere coupled to a split-ring resonator on a stripline\cite{kaur2016voltage}, and Zhang et al. obtained 32.85 MHz for a YIG cylinder/SRR stripline system\cite{zhang2017spin}. In comparison, the coupling strengths achieved in our planar YIG/HRR hybrid structure are tunable between 62.6 MHz and 126.6 MHz, which are significantly higher than the typical values reported for millimetre-scale spherical or cylindrical YIG resonators. These results indicate that while spherical YIG resonators benefit from large spin ensembles and mode volumes, carefully engineered planar geometries such as the present YIG/HRR design not only enable stronger and tunable coupling but also offer distinct advantages in terms of compatibility with on-chip integration.

Experimental studies have validated many of these strategies. For example, Bhoi et al.\cite{bhoi2018stress} demonstrated strain-induced damping modulation in nanometer-thick YIG films grown by pulsed laser deposition, while others achieved damping control via nonlocal spin currents\cite{haidar2016controlling} and substrate orientation modifications\cite{gurjar2021control}. Notably, Zhang et al.\cite{Zhang2014} observed the Purcell effect in a 3D microwave cavity, where enhanced cavity photon decay was induced by coupling with the lossy magnon mode of a YIG sphere. In that work, losses were deliberately introduced by attaching iron filings to the sphere, thereby increasing scattering and absorption. This experimental evidence strongly supports our findings that the Purcell effect in hybrid photon-magnon systems can be realized through multiple strategies for engineering lossy magnetic materials.

\section{Conclusion}
In summary, this study establishes a theoretical framework for realizing the Purcell effect in photon–magnon coupled systems, identifying magnon damping as a central tunable parameter for controlling spontaneous emission and photon dissipation. Through numerical simulations and parameter-space mapping, we show how damping, photon loss, and coupling strength collectively determine the transition between coherent coupling and the Purcell regime, while spin-density scaling confirms the cooperative nature of the effect. The feasibility of tuning YIG damping via growth optimization, rare-earth doping, spin pumping, and substrate engineering underscores the experimental practicality of our approach. By integrating magnetic control with Purcell engineering in a planar photon–magnon system, we demonstrate tunable collective light–matter interactions, revealing transitions between strong coupling and the Purcell regime and providing design principles for scalable, chip-compatible hybrid architectures.

\begin{acknowledgments}
The work was supported by the Council of Science and Technology, Uttar Pradesh (CSTUP), (Project Id: 2470, CST, U.P. sanction No: CST/D-1520). B. Bhoi acknowledges support by the Science and Engineering Research Board (SERB) India- SRG/2023/001355. R. Singh acknowledges support from the Council of Science and Technology, Uttar Pradesh (CSTUP), (Project Id: 4482). S. Verma acknowledges Ministry of Education, Government of India for the Prime Minister's Research Fellowship (PMRF ID-1102628).
\end{acknowledgments}

\section*{Data Availability Statement}
The data that support the findings of this study are available within the article.

\section*{References}

%

\end{document}